\documentclass[copyright,creativecommons]{eptcs}
\providecommand{\event}{ICLP 2026}

\usepackage{iftex}

\ifpdf
  \usepackage{underscore}         
  \usepackage[T1]{fontenc}        
\else
  \usepackage{breakurl}           
\fi

\usepackage{xcolor}
\usepackage[breakable]{tcolorbox}
\usepackage{enumitem}
\usepackage{xspace}
\usepackage[nolist]{acronym}
\usepackage[caption=false,font=scriptsize]{subfig}
\usepackage{tikz}
\usetikzlibrary{positioning,shapes,backgrounds,calc}
\usepackage{calc}
\usepackage{longtable}
\usepackage{booktabs}
\usepackage{amsmath}

\pgfkeyssetvalue{/change colors/none}{red}
\pgfkeyssetvalue{/change colors/RB}{orange}
\pgfkeyssetvalue{/change colors/VMM}{green}

\newcommand{\changed}[2][none]{{\pgfkeysgetvalue{/change colors/#1}{\changecolor}\color{\changecolor}#2}}
\renewcommand{\changed}[2][none]{#2}

\newcommand{\prohelp}[0]{\textsc{ProHelp}\xspace}
\newcommand{\gitseed}[0]{\textsc{GitSEED}\xspace}
\newcommand{\gitlab}[0]{\textsc{GitLab}\xspace}

\acrodef{SUS}{System Usability Scale}
\acrodef{LLM}{Large Language Model}
\acrodef{AAT}{Automated Assessment Tool}
\acrodef{IDE}{Integrated Development Environment}
\acrodef{CI}{Continuous Integration}

\tcbset{colback=black!5!white,boxrule=1pt,colframe=black!40!white,size=title}

\definecolor{c1dark}{RGB}{140,81,10}
\definecolor{c2dark}{RGB}{1,102,94}
\definecolor{c1light}{RGB}{246,232,195}
\definecolor{c2light}{RGB}{199,234,229}

\title{Can Automated Feedback Turn Students\\into Happy Prologians?}

\author{Ricardo Brancas
\institute{INESC-ID / IST,\\Universidade de Lisboa,\\Portugal}
\and
Pedro Orvalho
\institute{Artificial Intelligence Research Institute,\\
Consejo Superior de Investigaciones Científicas,\\Barcelona, Catalonia, Spain}
\and
Carolina Carreira
\institute{Carnegie Mellon University,\\Pittsburgh, USA}
\institute{INESC-ID / IST, Universidade de Lisboa, Portugal}
\and
Vasco Manquinho
\institute{INESC-ID / IST,\\Universidade de Lisboa,\\Portugal}
\and
Ruben Martins
\institute{Carnegie Mellon University,\\Pittsburgh, USA}}

\begin{document}

\providecommand{\titlerunning}{Can Automated Feedback Turn Students into Happy Prologians?}
\providecommand{\authorrunning}{R. Brancas, P. Orvalho, C. Carreira, V. Manquinho \& R. Martins}
\providecommand{\copyrightholders}{Brancas, Orvalho, Carreira, Manquinho \& Martins}

\maketitle

\begin{abstract}

Providing personalized feedback is essential for effective learning, but delivering it promptly can be challenging in large-scale courses.
\changed[RB]{In this work, we present \prohelp, an automated assessment platform for Prolog built on top of the \gitseed framework, and we evaluate it through a survey of 144 students from a 365-student undergraduate logic programming course.
We assessed the perceived usefulness of seven types of automated feedback, including automatic testing, predicate scoring, syntax error highlighting, open choice point warnings, score rankings, solution type validation, and unknown predicate name suggestions.
Our results show that 74\% of students agreed the feedback helped increase their grade, and the system achieved a System Usability Scale score of 78.5~(grade~B+). Among the feedback types, automatic testing was ranked as the most useful, followed by open choice point warnings and predicate scoring, with statistically significant differences.
We found no significant effect of students' interest level, engagement with optional exercises, or use of large language models on their perception of feedback usefulness.
We also explore student preferences for future feedback features, finding a significant preference for showing the differences between generated and expected test outputs.}
\end{abstract}

\section{Introduction}
\label{sec:introduction}

Giving timely and formative feedback to students is very important for the learning process~\cite{DBLP:journals/jeric/KeuningJH19,feedback2}.
However, doing so in large courses is difficult~\cite {DBLP:journals/jeric/PaivaLF22} and essentially impossible in cases such as Massive Open Online Courses (MOOCs). Over the years, many automated assessment frameworks have been proposed to address this problem. Automated assessment tools such as \textsc{Mooshak}/\textsc{Enki}~\cite{DBLP:conf/iticse/PaivaLQ16a} or \gitseed~\cite{DBLP:conf/sigcse/OrvalhoJM24}, among others~\cite{webcat08,ics23-seet-GradeStyle,sigcse17-submitty,geec16-codeOcean},
allow faculty to automatically test students' submissions and provide them with varying levels of feedback.

Many newer automated assessment systems focus on (1) making it more usable for students by targeting well-known interfaces such as {\sc Git}, and (2) making it easier to maintain and modify by faculty. Such tools, like \gitseed, allow faculty to fully customize the feedback returned to users. This makes it easy to leverage state-of-the-art software analysis and automated diagnosis tools, which in turn allow students to find and repair their mistakes more quickly.

However, despite these advances, significant challenges remain in the context of logic programming.
Novice students usually find it hard to debug logic programs. This is mainly due to the lack of traditional control and data flows they learn in imperative languages. In this work, we analyze these difficulties and students' behavior when learning Prolog.
To accomplish this, we created \prohelp, an automated assessment platform for Prolog built on top of \gitseed. \prohelp includes feedback features ranging from automated testing to typo-correction suggestions and validation of implementation techniques.

We designed and deployed a mixed-methods study with the students who interacted with \prohelp.  We used survey responses collected during a 9-week undergraduate course with 365 students. Overall, we analyzed 144 survey responses.

The main contributions of this work are:

\begin{itemize}
    \item Understanding which types of feedback implemented in \prohelp students find the most helpful;
    \item Gauging students' interest in more complex types of automated feedback they would like to see in the future.
\end{itemize}

Results show that the feedback provided to students was helpful and that the \prohelp automated assessment system is user-friendly. Students also showed great interest in proposed improvements and suggested some themselves.

This paper is organized as follows: in \autoref{sec:related-work} we present an overview of related work. Following that, in \autoref{sec:course}, we present the course in which this study was conducted, and in \autoref{sec:feedback-types}, we describe the automated assessment system and the types of feedback provided. Then, in \autoref{sec:methodology}, we present the methodology for our study, and in \autoref{sec:results}, we present the analysis and results. Finally, in \autoref{sec:discussion}, we present a discussion on the obtained results and conclude the paper.

\section{Related Work}
\label{sec:related-work}

\paragraph{\acfp{AAT}.} Automated assessment of programming assignments has a long history of more than 60 years of research~\cite{DBLP:journals/jeric/PaivaLF22}. Most \acp{AAT} assess programs by either comparing them with a known-correct solution or by executing a series of tests and checking their results~\cite{DBLP:journals/jeric/KeuningJH19}. A large number of current \acp{AAT} are based around \acl{IDE}-like web-interfaces with examples like \textsc{CodeOcean}~\cite{geec16-codeOcean}, \textsc{Mooshak}/\textsc{Enki}~\cite{DBLP:journals/spe/LealS03,DBLP:conf/iticse/PaivaLQ16a} and \textsc{Web-Cat}~\cite{webcat08}. Other platforms, like \textsc{ProgEdu}~\cite{ProgEdu}, GitHub Classroom\footnote{\url{https://classroom.github.com/}}, and \gitseed~\cite{DBLP:conf/sigcse/OrvalhoJM24} focus on Git, taking advantage of \ac{CI} actions to assess students' work as it is submitted to the repository. These tools can be easier for students to use because they are often already familiar with the Git workflow.

\paragraph{Automated Feedback for Logic Programming.} Many systems have been proposed for automated feedback, debugging, and repair of logic programming languages~\cite{le2011incom,DBLP:conf/iticse/MansouriGH98,DBLP:conf/issta/ThompsonS20,DBLP:conf/icst/BrancasMM25}. These tools face different challenges compared with other programming paradigms due to the declarative nature of the languages, lack of explicit control flow, and issues like non-determinism, non-monotonic reasoning, or non-termination~\cite{DBLP:conf/iticse/MansouriGH98}.
\textsc{INCOM}~\cite{le2011incom} introduces a two-stage tutoring model in which students first engage in task analysis, then implement Prolog predicates. This structured approach assists learners in bridging the gap between problem specification and algorithm development. Meanwhile, \textsc{PRAM}~\cite{DBLP:conf/iticse/MansouriGH98} focuses on the automatic assessment of Prolog programs by incorporating both static and dynamic metrics to evaluate code style, complexity, and dynamic correctness, thereby reducing instructor workload and facilitating the shift from procedural to declarative thinking.
Complementing these educational tools, \textsc{ProFL}~\cite{DBLP:conf/issta/ThompsonS20} applies fault localization techniques, particularly coverage-based and mutation-based approaches, to identify errors in Prolog programs effectively. Moreover, the \textsc{FormHe}~\cite{DBLP:conf/icst/BrancasMM25} system extends automated feedback to Answer Set Programming by combining traditional fault localization methods with modern strategies such as \ac{LLM}-based fault localization and repair and mutation-based program repair techniques.

One common element of earlier works on automated logic programming feedback is the use of structured, rule-based feedback, which can force students to solve problems in a certain way. Meanwhile, more recent approaches follow a test-based fault localization method, which helps students identify problems in their programs rather than guiding them in a particular direction. However, these newer tools lack comprehensive user studies to evaluate their performance. Our work addresses this research gap by directly engaging with the target audience of the tool -- the students -- and prioritizing their needs.

\section{Course Structure} \label{sec:course}

We evaluated students' behavior during a 9-week-long bachelor's level logic programming course. During this course, students learned classic logic reasoning, as well as logic programming through Prolog. Overall, students faced three different types of Prolog exercises during this course: a 5-part role-playing exercise to familiarize students with the programming language, 22 optional recitation exercises, and one graded and mandatory final project. The mandatory project was worth 50\% of the final grade, while the remaining 50\% was from other theoretical assessments. Next, we describe the Prolog exercises in more detail.

\begin{itemize}
    \item \textbf{Role-playing Exercise}
    \begin{itemize}
        \item Optional, not graded, online only
        \item 5 puzzles over 5 days
        \item A series of simple logic puzzles where students help a detective catch a group of criminals using logic programming;
    \end{itemize}
    \item \textbf{Recitation Exercises}
    \begin{itemize}
        \item Optional, not graded, partially solved in-class using pen and paper
        \item 22 exercises
        \item 4 Prolog exercise sheets grouped into the topics of: lists, arithmetic, negation, and higher-order functions;
    \end{itemize}
    \item \textbf{Project}
    \begin{itemize}
        \item Mandatory, graded
        \item 5 weeks duration
        \item Worth 50\% of the final course grade. The goal of this year's project was to create a solver for the logic puzzle \textit{star battle};
    \end{itemize}
\end{itemize}

\begin{figure}[tb]
    \centering
    \scalebox{0.9}{\begin{tikzpicture}[
week/.style={signal,draw,signal from=west,minimum width=9em},
c1tag/.style={draw=c1dark,line width=0.75pt,fill=c1light!50!white},
c2tag/.style={draw=c2dark,line width=0.75pt,fill=c2light!50!white},
]
    \node[text height=1.2ex] (e1) {\(\cdots\)};
    \node[week] (w3) [right=0.5em of e1] {Week 3};
    \node[week] (w4) [right=0em of w3] {Week 4};
    \node[week] (w5) [right=0em of w4] {Week 5};
    \node[text height=1.2ex] (e2) [right=0em of w5] {\(\cdots\)};
    \node[week,signal to=nowhere] (w9) [right=0.5em of e2] {Week 9};

    \coordinate (b0312) at ($(w3.south west)!5.5/7!(w3.south east)$);
    \coordinate (t0412) at ($(w3.north west)!6.5/7!(w3.north east)$);
    \coordinate (t0512) at ($(w4.north west)!0.5/7!(w4.north east)$);
    \coordinate (b0612) at ($(w4.south west)!1.5/7!(w4.south east)$);
    \coordinate (t0612) at ($(w4.north west)!1.5/7!(w4.north east)$);
    \coordinate (b0912) at ($(w4.south west)!4.5/7!(w4.south east)$);
    \coordinate (t0912) at ($(w4.north west)!4.5/7!(w4.north east)$);
    \coordinate (t1012) at ($(w4.north west)!5.5/7!(w4.north east)$);
    \coordinate (b1512) at ($(w5.south west)!3.5/7!(w5.south east)$);
    \coordinate (b1301) at ($(w9.south west)!4.5/7!(w9.south east)$);
    
    \draw[line width=1pt,color=c2dark] (b0312) -- +(0,-1em);
    \draw[line width=1pt,color=c2dark] (t0412) -- +(0,1em);
    \draw[line width=1pt,color=c2dark] (t0512) -- +(0,1em);
    \draw[line width=1pt,color=c2dark] (t0612) -- +(0,1em);
    \draw[line width=1pt,color=c1dark] (b0612) -- +(0,-4em);
    \draw[line width=1pt,color=c2dark] (b0612) -- +(0,-1em);
    \draw[line width=1pt,color=c2dark] (t0912) -- +(0,1em);
    \draw[line width=1pt,color=c1dark] (b0912) -- +(0,-2em);
    \draw[line width=1pt,color=c2dark] (t1012) -- +(0,1em);
    \draw[line width=1pt,color=c2dark] (b1512) -- +(0,-1em);
    \draw[line width=1pt,color=c1dark] (b1301) -- +(0,-1em);

    \draw[line width=1pt,color=c2dark] ($(t0412)+(-0.5pt,1em)$) -- ($(t1012)+(0.5pt,1em)$);
    \draw[line width=1pt,color=c2dark] ($(b0312)+(-0.5pt,-1em)$) -- ($(b1512)+(0.5pt,-1em)$);

    \coordinate (b0612end) at ($(b0612)-(0,4em)$);
    \coordinate (b0912end) at ($(b0912)-(0,2em)$);
    \coordinate (b1301end) at ($(b1301)-(0,1em)$);
    
    \node[c1tag,align=center,below right = 0em and -2em,at=(b0612end)] {\footnotesize Project Statement Published};

    \node[c1tag,align=center,below right = 0em and -2em,at=(b0912end)] {\footnotesize Project Repositories Published};

    \node[c1tag,align=center,below left = 0em and -2em,at=(b1301end)] {\footnotesize Project Deadline};

    \draw[line width=1pt,color=c2dark] ($(b0312)+(0.5em,-1em)$) -- ($(b0312)+(0.5em,-6em)$);

    \coordinate (b0312end) at ($(b0312)+(0.5em,-6em)$);

    \node[c2tag,align=center,below right = 0em and -2em,at=(b0312end)] {\footnotesize Recitation Exercises Published};

    \coordinate (t0412mid) at ($(w3.north west)!1!(w3.north east)+(0,1em)$);
    \coordinate (t0412end) at ($(w3.north west)!1!(w3.north east)+(0,2em)$);
    
    \draw[line width=1pt,color=c2dark] (t0412mid) -- (t0412end);

    \node[c2tag,align=center,above right = 0em and -2em,at=(t0412end)] {\footnotesize Role-playing Exercise Published};
    
\end{tikzpicture}}
    \caption{Course exercises timeline.}
    \label{fig:timeline}
\end{figure}

In \autoref{fig:timeline}, we show the timeline of the different exercises solved by the students. During the first weeks, the students learned logic fundamentals. Then, during week 3, the students started learning Prolog, with the optional exercise sets being released during weeks 3, 4, and 5. Meanwhile, the project statement was released at the beginning of week 4 with the deadline at the end of week 9.

\section{\prohelp: Prolog Automated Assessment Platform} \label{sec:feedback-types}

\begin{figure}[tb]
    \centering
    \includegraphics[width=0.95\linewidth]{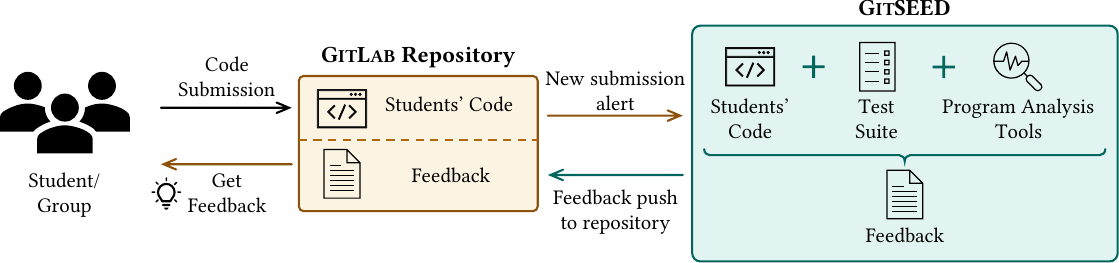}
    \caption{Simplified overview of \gitseed's operation scheme.}
    \label{fig:gitseed}
\end{figure}

The submission and evaluation platform used in this course was \prohelp, an extension built on top of the open-source automated assessment framework \gitseed~\cite{DBLP:conf/sigcse/OrvalhoJM24}. \gitseed provides students with personalized feedback on programming assignments while they learn \textsc{Git} fundamentals. Unlike traditional assessment platforms, \gitseed operates within \gitlab{}’s \ac{CI} workflow, eliminating the need for students to learn new interfaces.
As \autoref{fig:gitseed} shows, when students submit their work to \gitlab, \gitseed automatically evaluates it using a predefined test suite and program analysis tools specified by the faculty. The evaluation report is then pushed directly to the students’ git repositories, ensuring immediate access to personalized feedback.

\begin{figure}[tb]
    \centering
    \subfloat[Automatic testing\label{fig:screen-automatic-testing}]{%
        \includegraphics[width=0.24\linewidth]{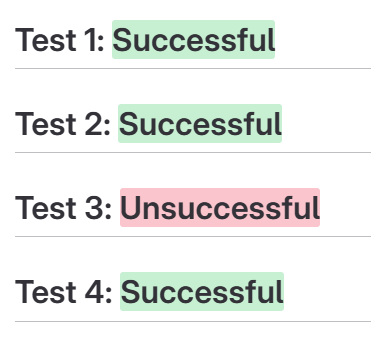}
    }\hfill
    \subfloat[Predicate Scoring\label{fig:screen-predicate-scoring}]{%
        \includegraphics[width=0.29\linewidth]{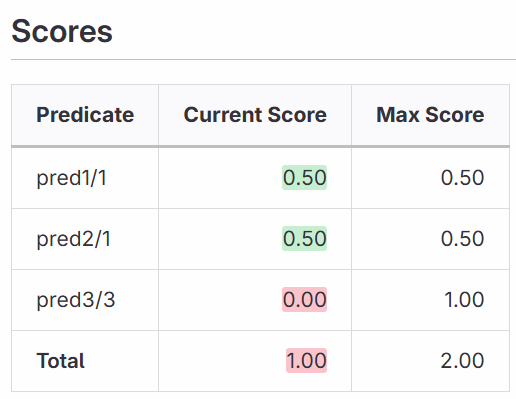}
    }\hfill
    \subfloat[Syntax error highlighting\label{fig:screen-syntax-error}]{%
        \includegraphics[width=0.39\linewidth]{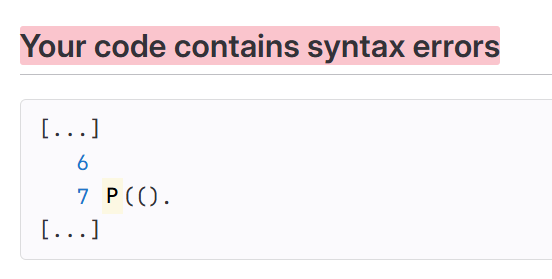}
    }

    \subfloat[Open choice points\label{fig:screen-open-choice-points}]{%
        \includegraphics[width=0.29\linewidth]{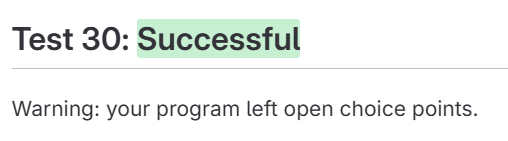}
    }\hfill
    \subfloat[Score rankings\label{fig:screen-score-rankings}]{%
        \includegraphics[width=0.69\linewidth]{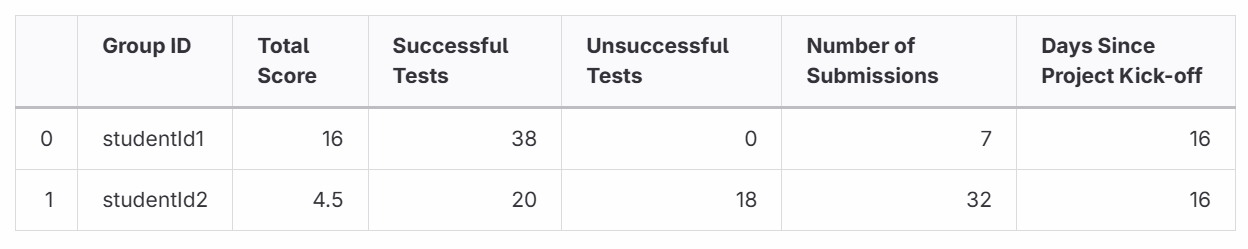}
    }
    
    \subfloat[Solution type validation\label{fig:screen-solution-type}]{%
        \includegraphics[width=0.85\linewidth]{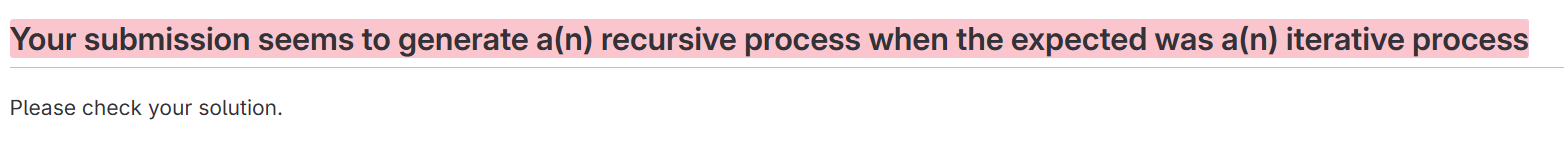}
    }

    \subfloat[Unknown predicate name suggestions\label{fig:screen-unknown-predicate}]{%
        \includegraphics[width=0.5\linewidth]{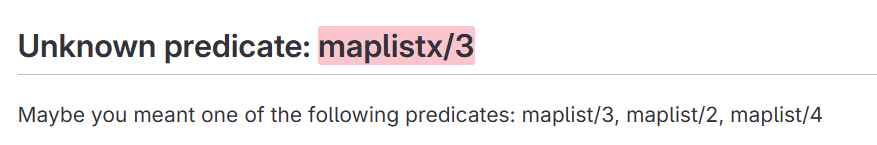}
    }
    
    \caption{Screenshots of the different types of student feedback available in \prohelp.}
    \label{fig:screenshots}
\end{figure}

\subsection{Automated Feedback}

The minimum configuration of \gitseed simply runs a test suite on the student's submission and reports the results as feedback. However, as shown in \autoref{fig:gitseed}, \gitseed can be extended with custom program-analysis tools to provide students with more personalized feedback. \changed[RB]{Building on \gitseed's standard automatic testing~(a) and score rankings~(e), we created \prohelp to implement five additional types of feedback specific to the Prolog context.} Below, we describe each of the 7 types of automated feedback, along with an example for each one, shown in \autoref{fig:screenshots}.

\begin{enumerate}[label=\alph*),itemsep=0.7ex]
    \item \textbf{Automatic testing:}\hspace{0.3em} This is the basic feature consisting of running the test suite and reporting the results. The faculty can decide to show the body of failing test cases or not.
    
    \item \textbf{Predicate scoring:}\hspace{0.3em} For graded assignments, each predicate \(p\) that the students need to implement gets assigned X points by the faculty. If a student passes the tests for \(p\), then they get X points.
    
    \item \textbf{Syntax error highlighting:}\hspace{0.3em} \prohelp highlights snippets of code with syntax errors, which makes them easier to locate compared with the line and column coordinates provided by the Prolog~interpreter.
    
    \item \textbf{Open choice points warning:}\hspace{0.3em} Prolog allows enumerating multiple answers for a given query. However, students often forget this fact. This can result in the first answer being correct with respect to the specification, but one of the following answers being incorrect. When a predicate terminates with an open choice point (indicating there are more possible answers), we warn students so they can verify whether this is intended behavior.
    
    \item \textbf{Score rankings:}\hspace{0.3em} Each assignment has a global leaderboard that shows the scores and number of passed tests for each student. This incentivizes some students to compete and solve the most exercises as early as possible.
    
    \item \textbf{Solution type validation:}\hspace{0.3em} Some of the assignments ask students to solve problems using specific techniques (for example, using recursion). In these assignments, we use a heuristic to determine if the student's program is using the correct method or not. The heuristic inspects the call dependencies of the main predicate: it is classified as \textit{functional} if it relies on built-in higher-order predicates (such as \texttt{maplist} or \texttt{foldl}), \textit{recursive} if it depends on itself, and \textit{iterative} if it delegates the work to an auxiliary predicate of higher arity (typically carrying an accumulator).
    
    \item \textbf{Unknown predicate name suggestions:}\hspace{0.3em} When a student uses a predicate that is undefined, \prohelp suggests other predicates with similar names if they exist.
    
\end{enumerate}

\subsection{Future Directions} \label{sec:future-feedback}

One of the goals of this work is to gauge students' opinions on the feedback features already implemented and decide which ones to prioritize for future work. Planned features include several AI-based procedures using formal methods targeted to the specific context of logic programming.

\begin{itemize}[itemsep=0.7ex]
    \item \textbf{Fault localization:}\hspace{0.3em} Identifying what section of the program contains bugs could help students narrow their focus. The granularity of the bug location could also be controlled by the faculty, such as identifying a specific predicate or a specific set of lines.

    \item \textbf{Bug fix suggestions:}\hspace{0.3em} The bug identification can be taken one step further and also include a description of the type of bug and/or a hint on how to fix it. One way to accomplish this is to use automated program repair to find a bug fix, then combine the buggy and corrected programs to generate a hint.
    
    \item \textbf{Infinite loop identification:}\hspace{0.3em} Prolog programs can suffer from infinite loops due to the depth-first search algorithm used by default in many Prolog interpreters. Helping students identify which sections of the program are responsible for the infinite loop behavior can be a helpful tool for~students.

    \item \textbf{Test output differences:}\hspace{0.3em} The test suites in \prohelp are composed of unit tests implemented through \texttt{plunit}.
    Almost always, these tests consist of a series of predicate calls followed by an assertion that compares a value with the expected response. It could be useful for students to provide more information when these assertions fail, particularly showing the difference between generated and expected values.
\end{itemize}

\section{Methodology}
\label{sec:methodology}

\prohelp was deployed in a BSc-level class on logic reasoning and programming with 365 students at Universidade de Lisboa, Portugal, during the 2024 academic year. Students were given automated feedback on 3 different types of Prolog assignments: a role-playing exercise, recitation exercises, and a final project.

To understand the strengths and shortcomings of \prohelp, we performed a post-experience survey. The survey, which is replicated in \autoref{app:survey}, had an estimated duration of 15 minutes and was performed using EUSurvey\footnote{\url{https://ec.europa.eu/eusurvey}} during the two weeks following the last programming assignment.

Of the 312 students who interacted with \prohelp during the course, we obtained 148 survey responses (47\%), of which we dropped 4 due to failed attention checks~\cite{Oppenheimer2009InstructionalMC}.
Of the 144 valid responses, 28 respondents identified as female (19\%), 111 as male (77\%), 1 as other (<1\%), and 4 preferred not to disclose (3\%). While we have no information on the gender distribution of students who took this class, the official distribution of enrolled students in this BSc degree is 83\% male and 17\% female. This provides some confidence that our sample is representative of the overall course population.

In this work, we plan to answer the following three research questions:

\begin{enumerate}[label=\textbf{RQ\arabic*.}]
    \item How valuable did students find the feedback, and which aspects of it were the most useful?
    \item Does student interest, engagement or \acs*{LLM}-usage affect how helpful they found the feedback?
    \item What features do students believe would be most beneficial in the future?
\end{enumerate}

\subsection{Data Analysis}

Part of our study focuses on analyzing 3- and 5-point Likert scales regarding the usefulness of different aspects of the feedback received. For RQ1 and RQ3, we are interested in how a student's different answers compare to one another (the ratings for the different types of feedback). To accomplish this, we use the Friedman rank sum test~\cite{Friedman1937TheUO} to check for any statistically significant differences between the feedback types. In cases where there are such differences, we employ the post-hoc pairwise Durbin-Conover test~\cite{conover1999practical} with the Holm correction method~\cite{Holm1979ASS}, in order to identify which pairs of feedback types have statistically significant differences in usefulness.

In RQ2, we are interested in how different sub-groups of students rated the usefulness of \prohelp~(on a 5-point Likert scale). To accomplish this, we use different tests depending on the nature of the independent variable. Students' interest was measured on a scale from 1 to 5 (an ordinal variable) and, as such, we use Spearman's rank correlation~\cite{spearman1904} to compare it with the usefulness. Meanwhile, students' engagement and usage of \acp{LLM} were both measured through categorical variables. We use the Kruskal-Wallis test~\cite{hollander2013nonparametric} for engagement, since it has more than 2 categories, and the Mann-Whitney~\(U\) test~\cite{10.1214/aoms/1177730491} for LLM usage, since it has only 2 categories.
The significance level for all tests was set at \(0.05\).

The survey also included two open-ended questions about future feedback suggestions and general opinions on the system. The answers were coded according to a unified code-book for Feedback and Improvements. The code-book was created by the first author. The answers were coded independently by the first and second authors, with any differences reconciled through a subsequent discussion.

\subsection{Ethics and Privacy}

All students had access to the same system and the same types of feedback. This ensured that no group of students had any advantage over the rest. Participation in the survey was voluntary, unremunerated, and all participants provided informed consent for data collection and processing before the survey. Participants were recruited through a course announcement after the last assignment. No identifiable personal data was collected. At Universidade de Lisboa, where the course was held and the survey was conducted, studies with these characteristics do not need to be submitted to the Ethics Committee (as confirmed by the Chair of the Ethics Committee). Co-authors from institutions other than Universidade de Lisboa had access only to aggregate statistics, not to individual responses. We have completed the European Commission's ethics self-assessment checklists and comply with all the guidelines.

\section{Results}
\label{sec:results}

This work aims to understand the advantages and limitations of the \prohelp system and explore what types of feedback students would like to have in the future. The statistical analysis and plots presented in this section were created using R version 4.4.2~\cite{Rlanguage}, statsExpressions~\cite{DBLP:journals/jossw/Patil21a}, ggstats~\cite{ggstats} and ggsignif~\cite{ggsignif}.

\begin{figure}[tb]
    \centering
    \includegraphics[width=0.9\textwidth]{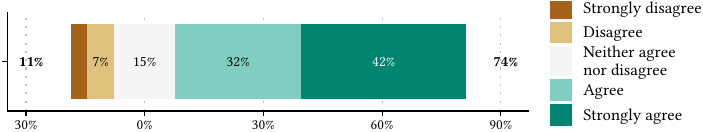}
    \caption{Agreement distribution for the statement ``I think the automated feedback helped increase my grade in this course.''.}
    \label{fig:increased-grade}
\end{figure}

\subsection{Usefulness and usability of the feedback (RQ1)}

To gauge if \prohelp was useful to students, we asked them if they agreed with the statement ``I think the automated feedback
helped increase my grade in this course'' using a 5-point Likert scale. Figure~\ref{fig:increased-grade} shows that 74\% of student responses agreed that the provided feedback was useful, while 11\% disagreed and 15\% neither agreed nor disagreed. This indicates that \prohelp was helpful for students.
We also performed a usability analysis through the \ac{SUS} questionnaire~\cite{Brooke1996SUSA}. The average \ac{SUS} score among our respondents was 78.5 which corresponds to a B+ grade according to the curved grading interpretation by Sauro and Lewis~\cite{sauro2016quantifying}. While there is room for improvement, the results show that our system is fairly usable in its current state.

\begin{figure}[tb]
    \centering
    \includegraphics[width=0.9\textwidth]{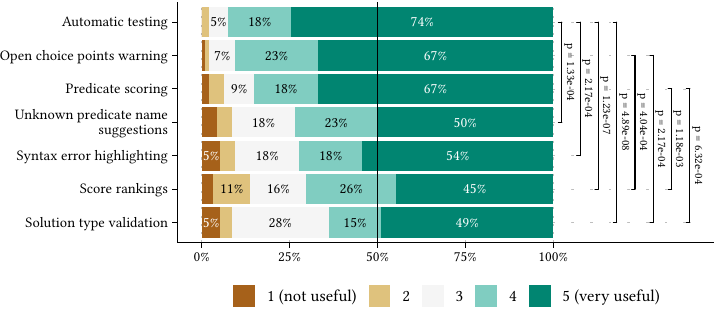}
    \caption{Usefulness rating of the different types of feedback. The median response for each type is shown by the vertical line. The right side of the figure shows pairs of feedback types for which the Durbin-Conover post-hoc test indicates statistically significant differences in ratings. For example, there is a statistically significant difference between the responses for ``Automatic testing'' and ``Score rankings'' with \(p = 1.23\mathrm{e}\text{-}07\).}
    \label{fig:feedback-types}
\end{figure}

This study also aims to understand if any type of feedback stood out as particularly more useful than the others. Hence, we asked the students to rank each of the 7 types of feedback in Section~\ref{sec:feedback-types} on a scale from 1 (not useful) to 5 (very useful). Figure~\ref{fig:feedback-types} presents the answer distribution to these questions. Most students ranked all 7 types of feedback as useful, with all of the medians being \(\ge 4\).

Since not all students interacted with all forms of feedback and the Friedman test requires a full block design, we excluded students who did not answer all 7 questions, resulting in 94 samples. The Friedman test allows us to reject the null hypothesis of there being no significant difference between the different types of feedback (\(\chi^2(6) = 58.66,\; p = 8.436 \times 10^{-11},\; \hat{W}_{Kendall}=0.10\)). As such, to discover which types of feedback were significantly more useful, we used the post-hoc pairwise Durbin-Conover test. The right side of Figure~\ref{fig:feedback-types} shows the pairs of feedback types for which there is a significant difference in median ranks, along with the respective p-values.
These statistically significant tests show that ``Automatic testing'' was the most helpful type of feedback, being more useful than 4 other types, and that ``Open choice points warning'' and ``Predicate scoring'' were more useful than 2 other types each. Meanwhile, the least helpful types of feedback were ``Score rankings'' and ``Solution type validation'', with each being less useful than 3 other types.

\begin{tcolorbox}
    \textbf{A1.} Most students consider that the provided feedback was helpful, and the system has a usability score of B+. The most useful feedback were ``Automatic testing'', ``Open choice points warning'' and ``Predicate scoring'', while the least useful were ``Score ranking'' and ``Solution type validation''. These conclusions are supported by statistically significant differences.
\end{tcolorbox}

\subsection{Impact of interest, engagement, and LLM use (RQ2)}

Another goal of this study is to determine if students' interest, engagement with optional exercises, and having used \acp{LLM} (or not) had any effect on their perception of the usefulness of the feedback.

\subsubsection{Interest}

\begin{figure}[tb]
    \centering
    \includegraphics[width=0.9\textwidth]{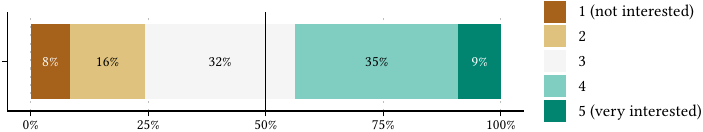}
    \caption{Answer distribution for the question ``How interested are you in the topics of logic and logic programming?''. The median response is shown by the black vertical line.}
    \label{fig:interest}
\end{figure}

We asked students to rate their interest in logic and logic programming on a scale from 1 to 5. The answers, summarized in Figure~\ref{fig:interest}, show a fairly normal distribution, with most students having some interest in the topics and a small number showing either no interest or a high level of interest.

\begin{figure}[tb]
    \centering
    \includegraphics[width=0.9\textwidth]{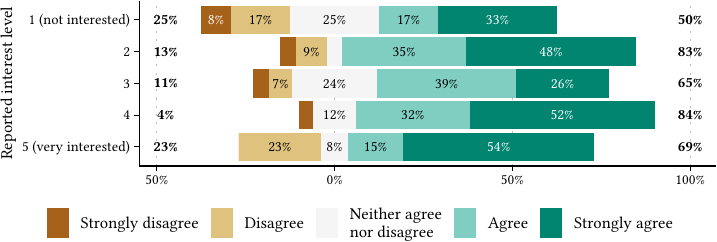}
    \caption{Effect of interest in the agreement distribution for the statement ``I think the automated feedback helped increase my grade in this course.''.}
    \label{fig:interest-effect}
\end{figure}

To test our hypothesis that more interested students take better advantage of automated feedback, we compared students' interest with the level of agreement with the statement ``I think the automated feedback helped increase my grade in this course''.  
Figure~\ref{fig:interest-effect} shows the agreement responses grouped by interest level.
A Spearman's rank correlation analysis does not allow us to reject the null hypothesis, which means there is no statistically significant relationship between students' interest level and how beneficial they found the feedback received (\(S = 4.26\times 10^{5},\; p = 0.08,\; \hat{\rho} = 0.14\)).

\subsubsection{Engagement}

\begin{figure}[tb]
    \centering
    \includegraphics[width=0.9\textwidth]{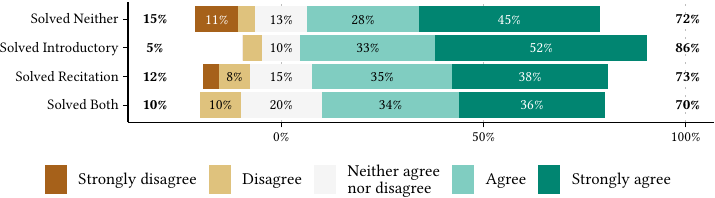}
    \caption{Effect of engagement in the agreement distribution for the statement ``I think the automated feedback helped increase my grade in this course.''.}
    \label{fig:engagement-effect}
\end{figure}

Students had the opportunity to use \prohelp with two types of optional exercises: a set of 5 introductory exercises and a set of 22 recitation exercises. Considering the 144 respondents to our survey, 47 did not solve any optional exercises, 50 solved at least one exercise from both optional sets, 21 solved only exercises from the introductory set, and 26 solved only exercises from the recitation set.
Our hypothesis is that more engaged students (ones who solved optional exercises) would have taken better advantage of \prohelp and thus found it more useful.
Figure~\ref{fig:engagement-effect} shows the relationship between the 4 engagement groups and the agreement levels with the statement ``I think the automated feedback helped increase my grade in this course''. To test our hypothesis, we used a Kruskal-Wallis test, which shows no evidence of there being a statistical difference between the four groups (\(\chi^2(3) = 2.19,\; p = 0.53,\; \hat{\epsilon}^2=0.02\)).

\subsubsection{Usage of LLMs}

\begin{table}[tb]
    \centering
    \small
    \begin{tabular}{lrrrrrrrr}
        \toprule
        \textbf{LLM Tool} & {ChatGPT} & {Copilot} & {Claude} & {Gemini} & {Blackbox} & {Deepseek} & {Perplexity} & {ZZZ Code} \\ \midrule
        \textbf{Count} & 82 & 16 & 5 & 2 & 1 & 1 & 1 & 1 \\
        \bottomrule
    \end{tabular}
    \caption{Number of students that reported using each LLM tool.}
    \label{tab:llm-usage}
\end{table}

\begin{figure}[tb]
    \centering
    \includegraphics[width=0.9\textwidth]{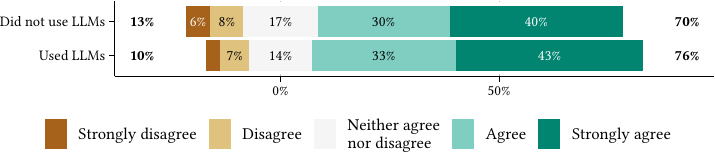}
    \caption{Effect of \acp{LLM} usage  in the agreement distribution for the statement ``I think the automated feedback helped increase my grade in this course.''.}
    \label{fig:llms-effect}
\end{figure}

Students were allowed to use \acfp{LLM} during this course, as long as they disclosed that usage. Considering the 144 responses to our survey, 91 reported using \acp{LLM} when solving assignments, while 53 reported that they did not use \acp{LLM}. \autoref{tab:llm-usage} shows the number of students who reported using each \ac{LLM} tool/platform, with the most popular ones being ChatGPT and Copilot. Furthermore, 21 students reported using more than one \ac{LLM}. We were interested in whether using \acp{LLM} affects how students perceive the usefulness of \prohelp (for example, students who used \acp{LLM} might have found \prohelp less useful because they already had personalized feedback from the \ac{LLM}).
In Figure~\ref{fig:llms-effect}, we show the usefulness ratings for students who used and did not use \acp{LLM}. According to the Mann-Whitney \(U\) test, there is no statistical difference between the two groups (\(W = 2261.50,\; p = 0.51,\; r = -0.06\)).

\begin{tcolorbox}
    \textbf{A2.} There is no evidence that interest level, having used the platform for optional exercises, or having used \acp{LLM}, affects how helpful students found \prohelp overall.
\end{tcolorbox}

\begin{figure}[tb]
    \centering
    \includegraphics[width=0.9\textwidth]{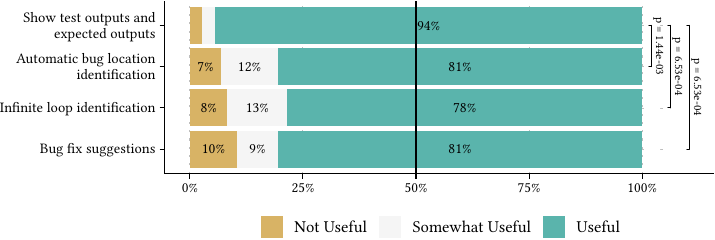}
    \caption{Usefulness ratings for the different types of feedback not yet implemented in \prohelp. The median response for each type of feedback is shown by the black vertical line. The right side of the figure shows pairs of feedback types for which the Durbin-Conover post-hoc test indicates statistically significant differences in ratings.}
    \label{fig:future-feedback-types}
\end{figure}

\subsection{Future directions (RQ3)}

Our last goal is to better understand which features and improvements to the feedback students would most like to see in the future, while also analyzing which aspects of the course students did not like. In the survey, we asked students to rate the 4 types of (not yet implemented) feedback described in Section~\ref{sec:future-feedback} between Not Useful, Somewhat Useful and Useful. 
Figure~\ref{fig:future-feedback-types} shows the students' answers, where at least 75\% of students rated each of the proposed feedback types as Useful. These responses hint that these are good goals to work towards.

To check if any type of feedback had a significant preference over another, we used a Friedman test, which rejects the null hypothesis (\(\chi^2(3) = 20.74,\; p = 1.19 \times 10^{-4},\; \hat{W}_{Kendall}=0.05\)). Next, to check which types of feedback students think will be more useful, we used the post-hoc pairwise Durbin-Conover test. The right side of Figure~\ref{fig:future-feedback-types} shows the pairs of feedback types that showed statistical significance.
Students rated ``Showing the differences between generated and expected output'' as significantly more useful than the other 3 types of feedback.

At the end of the survey, we asked students to suggest other types of feedback they would like to see in the future and to provide general comments/criticisms.
One common criticism among students was that the unit tests were not specific enough, i.e., each test covered too much functionality. For example, P-31 said: ``The project tests were not exhaustive enough.''. This left students with little information when tests failed and highlights the importance of faculty creating good unit tests that assess each functionality in isolation. There were also other comments regarding the course organization and grading method, which we have relayed to the faculty.

One interesting suggestion from some students concerns the qualitative component of the project grade, which is manually assigned and encompasses characteristics such as code readability, organization, and documentation. Students expressed a need for an estimate of this qualitative component in the automated feedback, even if the final score remained hand-assigned by faculty.

\begin{tcolorbox}
    \textbf{A3.} Although students found all proposed types of feedback useful, there is a significant preference for the ``Show the outputs and expected outputs'' compared with the remaining feedback types. 
    Students would like an automated estimate for the qualitative component of their projects.
\end{tcolorbox}

\section{Discussion and Conclusion}
\label{sec:discussion}

This work analyzes students' behaviors during a 9-week-long undergraduate class on logic programming. Specifically, we study how they interacted with an automated assessment platform for Prolog code. This platform, named \prohelp, included several types of automated feedback to help students find and fix their bugs more easily.
Our main contribution is the analysis of a student survey with 144 valid responses, which gathers insights into students' preferences and complaints regarding \prohelp.

Survey analysis indicates that users found the feedback received useful and the \prohelp interface was user-friendly. For instance, P-101 said \emph{``I loved the assessment system. It was very efficient and helpful. From what I have heard from previous years, it was a huge improvement''}. The components of \prohelp that students most appreciated include automatic testing, warnings about open choice points, and scoring for each required predicate. Students also displayed great interest in the more advanced types of feedback we proposed and suggested some themselves. These findings will be valuable for future research, enabling educators and developers to focus on the most impactful types of feedback.

Results show no statistically significant relationship between students’ perception of feedback usefulness and their interest, engagement, or usage of \acp{LLM}. Even so, there is a small but not statistically significant correlation (\(p=0.08\)) between students' interest and their usefulness rating, suggesting that further research -- with a larger sample or a different methodology -- is warranted. The findings on \ac{LLM} usage are particularly intriguing, and we propose several possible explanations for it: (1) students under-reported \ac{LLM}-usage due to its negative connotation, (2) the feedback received in \prohelp is orthogonal to the questions students usually ask \acp{LLM}, or (3) \acp{LLM} are not (yet) very useful for Prolog. Further investigation is needed to determine which, if any, of these factors apply.

\changed[RB]{Overall, our findings demonstrate that automated feedback is a valuable resource for students learning Prolog, and that even simple feedback mechanisms, when well integrated into familiar workflows, can have a meaningful impact on the learning experience. These insights can guide educators and tool developers in prioritizing the types of feedback that students find most impactful, ultimately contributing to more effective logic programming education.}

\section*{Acknowledgments}
This work supported by Portuguese national funds through FCT, under project 2023.14280.PEX (DOI: 10.54499/\-2023.14280.PEX). This work was also supported by grant PID2022-139835NB-C21 funded by MCIN/\-AEI/\-10.13039/\-501100011033 and by ERDF, EU; and by HORIZON-MSCA-2025-PF, project 101269051 — Sherlock4Py funded by REA, EU. This work was additionally supported by the Carnegie Mellon University Portugal Program and FCT under grants PRT/BD/152086/2021 (DOI: 10.54499/\-PRT/\-BD/\-152086/\-2021) and PRT/BD/153739/2021 (DOI: 10.54499/\-PRT/\-BD/\-153739/\-2021). This work was also partially supported by the National Science Foundation (NSF) under Award CCF2427581 and DARPA under Agreement FA8750-24-9-1000.

\bibliographystyle{eptcs}
\bibliography{bibliography}

\begin{thebibliography}{10}
\providecommand{\bibitemdeclare}[2]{}
\providecommand{\surnamestart}{}
\providecommand{\surnameend}{}
\providecommand{\urlprefix}{Available at }
\providecommand{\url}[1]{\texttt{#1}}
\providecommand{\href}[2]{\texttt{#2}}
\providecommand{\urlalt}[2]{\href{#1}{#2}}
\providecommand{\doi}[1]{doi:\urlalt{https://doi.org/#1}{#1}}
\providecommand{\eprint}[1]{arXiv:\urlalt{https://arxiv.org/abs/#1}{#1}}
\providecommand{\bibinfo}[2]{#2}

\bibitemdeclare{inproceedings}{DBLP:conf/icst/BrancasMM25}
\bibitem{DBLP:conf/icst/BrancasMM25}
\bibinfo{author}{Ricardo \surnamestart Brancas\surnameend},
  \bibinfo{author}{Vasco \surnamestart Manquinho\surnameend} \&
  \bibinfo{author}{Ruben \surnamestart Martins\surnameend}
  (\bibinfo{year}{2025}): \emph{\bibinfo{title}{Combining Logic and Large
  Language Models for Assisted Debugging and Repair of {ASP} Programs}}.
\newblock In: {\slshape \bibinfo{booktitle}{{IEEE} Conference on Software
  Testing, Verification and Validation, {ICST} 2025, Napoli, Italy, March 31 -
  April 4, 2025}}, \bibinfo{publisher}{{IEEE}}, pp. \bibinfo{pages}{646--657},
  \doi{10.1109/ICST62969.2025.10988950}.

\bibitemdeclare{incollection}{Brooke1996SUSA}
\bibitem{Brooke1996SUSA}
\bibinfo{author}{J.~B. \surnamestart Brooke\surnameend} (\bibinfo{year}{1996}):
  \emph{\bibinfo{title}{{SUS}: A '{Quick} and {Dirty}' {Usability} {Scale}}}.
\newblock In: {\slshape \bibinfo{booktitle}{Usability Evaluation In Industry}},
  \bibinfo{publisher}{CRC Press}, pp. \bibinfo{pages}{207--212}.

\bibitemdeclare{inproceedings}{ProgEdu}
\bibitem{ProgEdu}
\bibinfo{author}{Hsi-Min \surnamestart Chen\surnameend},
  \bibinfo{author}{Wei-Han \surnamestart Chen\surnameend},
  \bibinfo{author}{Nien-Lin \surnamestart Hsueh\surnameend},
  \bibinfo{author}{Chi-Chen \surnamestart Lee\surnameend} \&
  \bibinfo{author}{Chia-Hsiu \surnamestart Li\surnameend}
  (\bibinfo{year}{2017}): \emph{\bibinfo{title}{ProgEdu - an automatic
  assessment platform for programming courses}}.
\newblock In: {\slshape \bibinfo{booktitle}{2017 International Conference on
  Applied System Innovation (ICASI)}}, pp. \bibinfo{pages}{173--176},
  \doi{10.1109/ICASI.2017.7988376}.

\bibitemdeclare{book}{conover1999practical}
\bibitem{conover1999practical}
\bibinfo{author}{W.J. \surnamestart Conover\surnameend} (\bibinfo{year}{1999}):
  \emph{\bibinfo{title}{Practical Nonparametric Statistics}}.
\newblock \bibinfo{series}{Wiley Series in Probability and Statistics},
  \bibinfo{publisher}{Wiley}.

\bibitemdeclare{article}{ggsignif}
\bibitem{ggsignif}
\bibinfo{author}{Ahlmann-Eltze \surnamestart Constantin\surnameend} \&
  \bibinfo{author}{Indrajeet \surnamestart Patil\surnameend}
  (\bibinfo{year}{2021}): \emph{\bibinfo{title}{{ggsignif}: R Package for
  Displaying Significance Brackets for {'ggplot2'}}}.
\newblock {\slshape \bibinfo{journal}{PsyArxiv}}, \doi{10.31234/osf.io/7awm6}.
\newblock \urlprefix\url{https://psyarxiv.com/7awm6}.

\bibitemdeclare{inproceedings}{webcat08}
\bibitem{webcat08}
\bibinfo{author}{Stephen~H. \surnamestart Edwards\surnameend} \&
  \bibinfo{author}{Manuel~A. \surnamestart
  P{\'{e}}rez{-}Qui{\~{n}}ones\surnameend} (\bibinfo{year}{2008}):
  \emph{\bibinfo{title}{Web-CAT: automatically grading programming
  assignments}}.
\newblock In \bibinfo{editor}{June \surnamestart Amillo\surnameend},
  \bibinfo{editor}{Cary \surnamestart Laxer\surnameend},
  \bibinfo{editor}{Ernestina~Menasalvas \surnamestart Ruiz\surnameend} \&
  \bibinfo{editor}{Alison \surnamestart Young\surnameend}, editors: {\slshape
  \bibinfo{booktitle}{Proceedings of the 13th Annual {SIGCSE} Conference on
  Innovation and Technology in Computer Science Education, ITiCSE 2008, Madrid,
  Spain, June 30 - July 2, 2008}}, \bibinfo{publisher}{{ACM}},
  \bibinfo{address}{""}, p. \bibinfo{pages}{328},
  \doi{10.1145/1384271.1384371}.

\bibitemdeclare{article}{Friedman1937TheUO}
\bibitem{Friedman1937TheUO}
\bibinfo{author}{Milton \surnamestart Friedman\surnameend}
  (\bibinfo{year}{1937}): \emph{\bibinfo{title}{The Use of Ranks to Avoid the
  Assumption of Normality Implicit in the Analysis of Variance}}.
\newblock {\slshape \bibinfo{journal}{Journal of the American Statistical
  Association}} \bibinfo{volume}{32}, pp. \bibinfo{pages}{675--701}.
\newblock \urlprefix\url{https://doi.org/10.2307/2279372}.

\bibitemdeclare{book}{hollander2013nonparametric}
\bibitem{hollander2013nonparametric}
\bibinfo{author}{M.~\surnamestart Hollander\surnameend}, \bibinfo{author}{D.A.
  \surnamestart Wolfe\surnameend} \& \bibinfo{author}{E.~\surnamestart
  Chicken\surnameend} (\bibinfo{year}{2013}):
  \emph{\bibinfo{title}{Nonparametric Statistical Methods}}.
\newblock \bibinfo{series}{Wiley Series in Probability and Statistics},
  \bibinfo{publisher}{Wiley}.
\newblock \urlprefix\url{https://doi.org/10.1002/9781119196037}.

\bibitemdeclare{article}{Holm1979ASS}
\bibitem{Holm1979ASS}
\bibinfo{author}{Sture \surnamestart Holm\surnameend} (\bibinfo{year}{1979}):
  \emph{\bibinfo{title}{A Simple Sequentially Rejective Multiple Test
  Procedure}}.
\newblock {\slshape \bibinfo{journal}{Scandinavian Journal of Statistics}}
  \bibinfo{volume}{6}, pp. \bibinfo{pages}{65--70}.
\newblock \urlprefix\url{https://www.jstor.org/stable/4615733}.

\bibitemdeclare{inproceedings}{ics23-seet-GradeStyle}
\bibitem{ics23-seet-GradeStyle}
\bibinfo{author}{Callum \surnamestart Iddon\surnameend},
  \bibinfo{author}{Nasser \surnamestart Giacaman\surnameend} \&
  \bibinfo{author}{Valerio \surnamestart Terragni\surnameend}
  (\bibinfo{year}{2023}): \emph{\bibinfo{title}{{GRADESTYLE:} GitHub-Integrated
  and Automated Assessment of Java Code Style}}.
\newblock In: {\slshape \bibinfo{booktitle}{45th {IEEE/ACM} International
  Conference on Software Engineering: Software Engineering Education and
  Training, SEET@ICSE 2023, Melbourne, Australia, May 14-20, 2023}},
  \bibinfo{publisher}{{IEEE}}, \bibinfo{address}{""}, pp.
  \bibinfo{pages}{192--197}, \doi{10.1109/ICSE-SEET58685.2023.00024}.

\bibitemdeclare{article}{DBLP:journals/jeric/KeuningJH19}
\bibitem{DBLP:journals/jeric/KeuningJH19}
\bibinfo{author}{Hieke \surnamestart Keuning\surnameend},
  \bibinfo{author}{Johan \surnamestart Jeuring\surnameend} \&
  \bibinfo{author}{Bastiaan \surnamestart Heeren\surnameend}
  (\bibinfo{year}{2019}): \emph{\bibinfo{title}{A Systematic Literature Review
  of Automated Feedback Generation for Programming Exercises}}.
\newblock {\slshape \bibinfo{journal}{{ACM} Trans. Comput. Educ.}}
  \bibinfo{volume}{19}(\bibinfo{number}{1}), pp. \bibinfo{pages}{3:1--3:43}.
\newblock \urlprefix\url{https://doi.org/10.1145/3231711}.

\bibitemdeclare{manual}{ggstats}
\bibitem{ggstats}
\bibinfo{author}{Joseph \surnamestart Larmarange\surnameend}
  (\bibinfo{year}{2025}): \emph{\bibinfo{title}{ggstats: Extension to 'ggplot2'
  for Plotting Stats}}.
\newblock \urlprefix\url{https://larmarange.github.io/ggstats/}.
\newblock \bibinfo{note}{R package version 0.8.0,
  https://github.com/larmarange/ggstats}.

\bibitemdeclare{inproceedings}{le2011incom}
\bibitem{le2011incom}
\bibinfo{author}{Nguyen-Thinh \surnamestart Le\surnameend} \&
  \bibinfo{author}{Niels \surnamestart Pinkwart\surnameend}
  (\bibinfo{year}{2011}): \emph{\bibinfo{title}{INCOM: A web-based homework
  coaching system for logic programming}}.
\newblock In: {\slshape \bibinfo{booktitle}{Conference on Cognition and
  Exploratory Learning in Digital Age}}, \bibinfo{organization}{Citeseer}, pp.
  \bibinfo{pages}{43--50}.

\bibitemdeclare{article}{DBLP:journals/spe/LealS03}
\bibitem{DBLP:journals/spe/LealS03}
\bibinfo{author}{Jos{\'{e}}~Paulo \surnamestart Leal\surnameend} \&
  \bibinfo{author}{Fernando M.~A. \surnamestart Silva\surnameend}
  (\bibinfo{year}{2003}): \emph{\bibinfo{title}{Mooshak: a Web-based multi-site
  programming contest system}}.
\newblock {\slshape \bibinfo{journal}{Softw. Pract. Exp.}}
  \bibinfo{volume}{33}(\bibinfo{number}{6}), pp. \bibinfo{pages}{567--581}.
\newblock \urlprefix\url{https://doi.org/10.1002/spe.522}.

\bibitemdeclare{article}{10.1214/aoms/1177730491}
\bibitem{10.1214/aoms/1177730491}
\bibinfo{author}{H.~B. \surnamestart Mann\surnameend} \& \bibinfo{author}{D.~R.
  \surnamestart Whitney\surnameend} (\bibinfo{year}{1947}):
  \emph{\bibinfo{title}{{On a Test of Whether one of Two Random Variables is
  Stochastically Larger than the Other}}}.
\newblock {\slshape \bibinfo{journal}{The Annals of Mathematical Statistics}}
  \bibinfo{volume}{18}(\bibinfo{number}{1}), pp. \bibinfo{pages}{50 -- 60}.
\newblock \urlprefix\url{https://doi.org/10.1214/aoms/1177730491}.

\bibitemdeclare{inproceedings}{DBLP:conf/iticse/MansouriGH98}
\bibitem{DBLP:conf/iticse/MansouriGH98}
\bibinfo{author}{Fatima~Z. \surnamestart Mansouri\surnameend},
  \bibinfo{author}{Cleveland~Augustine \surnamestart Gibbon\surnameend} \&
  \bibinfo{author}{Colin~A. \surnamestart Higgins\surnameend}
  (\bibinfo{year}{1998}): \emph{\bibinfo{title}{{PRAM:} prolog automatic
  marker}}.
\newblock In \bibinfo{editor}{Gordon \surnamestart Davies\surnameend} \&
  \bibinfo{editor}{M{\'{\i}}che{\'{a}}l \surnamestart
  {\'{O}}'h{\'{e}}igeartaigh\surnameend}, editors: {\slshape
  \bibinfo{booktitle}{Proceedings of the 6th Annual Conference on the Teaching
  of Computing and the 3rd Annual {SIGCSE} Conference on Innovation and
  Technology in Computer Science Education, ITiCSE 1998, Dublin City
  University, Ireland, 18-21 August 1998}}, \bibinfo{publisher}{{ACM}}, pp.
  \bibinfo{pages}{166--170}, \doi{10.1145/282991.283108}.

\bibitemdeclare{article}{Oppenheimer2009InstructionalMC}
\bibitem{Oppenheimer2009InstructionalMC}
\bibinfo{author}{Daniel~M. \surnamestart Oppenheimer\surnameend},
  \bibinfo{author}{Tom \surnamestart Meyvis\surnameend} \&
  \bibinfo{author}{Nicolas \surnamestart Davidenko\surnameend}
  (\bibinfo{year}{2009}): \emph{\bibinfo{title}{Instructional manipulation
  checks: Detecting satisficing to increase statistical power}}.
\newblock {\slshape \bibinfo{journal}{Journal of Experimental Social
  Psychology}} \bibinfo{volume}{45}(\bibinfo{number}{4}), pp.
  \bibinfo{pages}{867--872}.
\newblock \urlprefix\url{https://doi.org/10.1016/j.jesp.2009.03.009}.

\bibitemdeclare{inproceedings}{DBLP:conf/sigcse/OrvalhoJM24}
\bibitem{DBLP:conf/sigcse/OrvalhoJM24}
\bibinfo{author}{Pedro \surnamestart Orvalho\surnameend},
  \bibinfo{author}{Mikol{\'{a}}s \surnamestart Janota\surnameend} \&
  \bibinfo{author}{Vasco \surnamestart Manquinho\surnameend}
  (\bibinfo{year}{2024}): \emph{\bibinfo{title}{{GitSEED}: {A Git-backed
  Automated Assessment Tool for Software Engineering and Programming
  Education}}}.
\newblock In \bibinfo{editor}{Mohsen \surnamestart Dorodchi\surnameend},
  \bibinfo{editor}{Ming \surnamestart Zhang\surnameend} \&
  \bibinfo{editor}{Stephen \surnamestart Cooper\surnameend}, editors: {\slshape
  \bibinfo{booktitle}{Proceedings of the 2024 {ACM} Virtual Global Computing
  Education Conference V. 1, {SIGCSE} Virtual 2024, Virtual Event, NC, USA,
  December 5-8, 2024}}, \bibinfo{publisher}{{ACM}}.
\newblock \urlprefix\url{https://doi.org/10.1145/3649165.3690106}.

\bibitemdeclare{article}{DBLP:journals/jeric/PaivaLF22}
\bibitem{DBLP:journals/jeric/PaivaLF22}
\bibinfo{author}{Jos{\'{e}}~Carlos \surnamestart Paiva\surnameend},
  \bibinfo{author}{Jos{\'{e}}~Paulo \surnamestart Leal\surnameend} \&
  \bibinfo{author}{{\'{A}}lvaro \surnamestart Figueira\surnameend}
  (\bibinfo{year}{2022}): \emph{\bibinfo{title}{Automated Assessment in
  Computer Science Education: {A} State-of-the-Art Review}}.
\newblock {\slshape \bibinfo{journal}{{ACM} Trans. Comput. Educ.}}
  \bibinfo{volume}{22}(\bibinfo{number}{3}), pp. \bibinfo{pages}{34:1--34:40}.
\newblock \urlprefix\url{https://doi.org/10.1145/3513140}.

\bibitemdeclare{inproceedings}{DBLP:conf/iticse/PaivaLQ16a}
\bibitem{DBLP:conf/iticse/PaivaLQ16a}
\bibinfo{author}{Jos{\'{e}}~Carlos \surnamestart Paiva\surnameend},
  \bibinfo{author}{Jos{\'{e}}~Paulo \surnamestart Leal\surnameend} \&
  \bibinfo{author}{Ricardo Alexandre~Peixoto \surnamestart
  de~Queir{\'{o}}s\surnameend} (\bibinfo{year}{2016}):
  \emph{\bibinfo{title}{Enki: {A} Pedagogical Services Aggregator for Learning
  Programming Languages}}.
\newblock In \bibinfo{editor}{Alison \surnamestart Clear\surnameend},
  \bibinfo{editor}{Ernesto \surnamestart Cuadros{-}Vargas\surnameend},
  \bibinfo{editor}{Janet \surnamestart Carter\surnameend} \&
  \bibinfo{editor}{Yv{\'{a}}n \surnamestart T{\'{u}}pac\surnameend}, editors:
  {\slshape \bibinfo{booktitle}{Proceedings of the 2016 {ACM} Conference on
  Innovation and Technology in Computer Science Education, ITiCSE 2016,
  Arequipa, Peru, July 9-13, 2016}}, \bibinfo{publisher}{{ACM}}, pp.
  \bibinfo{pages}{332--337}.
\newblock \urlprefix\url{https://doi.org/10.1145/2899415.2899441}.

\bibitemdeclare{article}{DBLP:journals/jossw/Patil21a}
\bibitem{DBLP:journals/jossw/Patil21a}
\bibinfo{author}{Indrajeet \surnamestart Patil\surnameend}
  (\bibinfo{year}{2021}): \emph{\bibinfo{title}{statsExpressions: {R} Package
  for Tidy Dataframes and Expressions with Statistical Details}}.
\newblock {\slshape \bibinfo{journal}{J. Open Source Softw.}}
  \bibinfo{volume}{6}(\bibinfo{number}{61}), p. \bibinfo{pages}{3236}.
\newblock \urlprefix\url{https://doi.org/10.21105/joss.03236}.

\bibitemdeclare{inproceedings}{sigcse17-submitty}
\bibitem{sigcse17-submitty}
\bibinfo{author}{Matthew \surnamestart Peveler\surnameend},
  \bibinfo{author}{Jeramey \surnamestart Tyler\surnameend},
  \bibinfo{author}{Samuel \surnamestart Breese\surnameend},
  \bibinfo{author}{Barbara \surnamestart Cutler\surnameend} \&
  \bibinfo{author}{Ana~L. \surnamestart Milanova\surnameend}
  (\bibinfo{year}{2017}): \emph{\bibinfo{title}{Submitty: An Open Source,
  Highly-Configurable Platform for Grading of Programming Assignments (Abstract
  Only)}}.
\newblock In \bibinfo{editor}{Michael~E. \surnamestart Caspersen\surnameend},
  \bibinfo{editor}{Stephen~H. \surnamestart Edwards\surnameend},
  \bibinfo{editor}{Tiffany \surnamestart Barnes\surnameend} \&
  \bibinfo{editor}{Daniel~D. \surnamestart Garcia\surnameend}, editors:
  {\slshape \bibinfo{booktitle}{Proceedings of the 2017 {ACM} {SIGCSE}
  Technical Symposium on Computer Science Education, {SIGCSE} 2017, Seattle,
  WA, USA, March 8-11, 2017}}, \bibinfo{publisher}{{ACM}},
  \bibinfo{address}{""}, p. \bibinfo{pages}{641},
  \doi{10.1145/3017680.3022384}.

\bibitemdeclare{manual}{Rlanguage}
\bibitem{Rlanguage}
\bibinfo{author}{\surnamestart {R Core Team}\surnameend}
  (\bibinfo{year}{2024}): \emph{\bibinfo{title}{R: A Language and Environment
  for Statistical Computing}}.
\newblock \bibinfo{organization}{R Foundation for Statistical Computing},
  \bibinfo{address}{Vienna, Austria}.
\newblock \urlprefix\url{https://www.R-project.org/}.

\bibitemdeclare{book}{sauro2016quantifying}
\bibitem{sauro2016quantifying}
\bibinfo{author}{Jeff \surnamestart Sauro\surnameend} \&
  \bibinfo{author}{James~R \surnamestart Lewis\surnameend}
  (\bibinfo{year}{2016}): \emph{\bibinfo{title}{Quantifying the user
  experience: Practical statistics for user research}}.
\newblock \bibinfo{publisher}{Morgan Kaufmann}.
\newblock \urlprefix\url{https://doi.org/10.1016/C2010-0-65192-3}.

\bibitemdeclare{article}{spearman1904}
\bibitem{spearman1904}
\bibinfo{author}{Charles \surnamestart Spearman\surnameend}
  (\bibinfo{year}{1904}): \emph{\bibinfo{title}{The proof and measurement of
  association between two things}}.
\newblock {\slshape \bibinfo{journal}{The American Journal of Psychology}}
  \bibinfo{volume}{15}(\bibinfo{number}{1}), pp. \bibinfo{pages}{72--101}.
\newblock \urlprefix\url{https://doi.org/10.2307/1412159}.

\bibitemdeclare{inproceedings}{geec16-codeOcean}
\bibitem{geec16-codeOcean}
\bibinfo{author}{Thomas \surnamestart Staubitz\surnameend},
  \bibinfo{author}{Hauke \surnamestart Klement\surnameend},
  \bibinfo{author}{Ralf \surnamestart Teusner\surnameend}, \bibinfo{author}{Jan
  \surnamestart Renz\surnameend} \& \bibinfo{author}{Christoph \surnamestart
  Meinel\surnameend} (\bibinfo{year}{2016}): \emph{\bibinfo{title}{CodeOcean -
  {A} versatile platform for practical programming excercises in online
  environments}}.
\newblock In: {\slshape \bibinfo{booktitle}{2016 {IEEE} Global Engineering
  Education Conference, {EDUCON} 2016, Abu Dhabi, United Arab Emirates, April
  10-13, 2016}}, \bibinfo{publisher}{{IEEE}}, \bibinfo{address}{""}, pp.
  \bibinfo{pages}{314--323}, \doi{10.1109/EDUCON.2016.7474573}.

\bibitemdeclare{inproceedings}{DBLP:conf/issta/ThompsonS20}
\bibitem{DBLP:conf/issta/ThompsonS20}
\bibinfo{author}{George \surnamestart Thompson\surnameend} \&
  \bibinfo{author}{Allison~K. \surnamestart Sullivan\surnameend}
  (\bibinfo{year}{2020}): \emph{\bibinfo{title}{ProFL: a fault localization
  framework for Prolog}}.
\newblock In \bibinfo{editor}{Sarfraz \surnamestart Khurshid\surnameend} \&
  \bibinfo{editor}{Corina~S. \surnamestart Pasareanu\surnameend}, editors:
  {\slshape \bibinfo{booktitle}{{ISSTA} '20: 29th {ACM} {SIGSOFT} International
  Symposium on Software Testing and Analysis, Virtual Event, USA, July 18-22,
  2020}}, \bibinfo{publisher}{{ACM}}, pp. \bibinfo{pages}{561--564}.
\newblock \urlprefix\url{https://doi.org/10.1145/3395363.3404367}.

\bibitemdeclare{article}{feedback2}
\bibitem{feedback2}
\bibinfo{author}{Benedikt \surnamestart Wisniewski\surnameend},
  \bibinfo{author}{Klaus \surnamestart Zierer\surnameend} \&
  \bibinfo{author}{John \surnamestart Hattie\surnameend}
  (\bibinfo{year}{2020}): \emph{\bibinfo{title}{The power of feedback
  revisited: A meta-analysis of educational feedback research}}.
\newblock {\slshape \bibinfo{journal}{Frontiers in psychology}}
  \bibinfo{volume}{10}, p. \bibinfo{pages}{487662}.

\end{thebibliography}

\appendix

\section{Survey Protocol} \label{app:survey}

The following survey was designed to collect anonymous feedback from undegraduate students enrolled in the \textit{Logic Programming} course regarding their experience with the automated feedback system provided via GitLab. The survey was estimated to take approximately 10 to 15 minutes. Several questions were included to assess usability, usefulness, and pedagogical effectiveness of the feedback tools. Some questions functioned as attention checks. The questions and answers presented here were translated from Portuguese.

\subsection*{Section 1: Demographics and Background}

\begin{itemize}
  \item \textbf{What gender do you identify with?}
  \begin{itemize}
    \item Female
    \item Male
    \item Prefer not to say
    \item Prefer to self-describe
  \end{itemize}

  \item \textbf{What is your level of interest in the topics covered in the Programming Languages course? (Logic, Prolog)}
  
  Rating scale from 1 (not interested) to 5 (very interested)

  \item \textbf{During this course, did you experience difficulties using Git/GitLab? Select all that apply:}
  \begin{itemize}
    \item Merge conflicts
    \item SSH key problems
    \item Problems interacting with VS Code
    \item I had no problems
    \item Other
  \end{itemize}
\end{itemize}

\subsection*{Section 2: Automated Feedback in Prolog}

Students were reminded of the types of automated feedback they encountered on GitLab:

\begin{itemize}
  \item Automatic test verification
  \item Predicate scoring
  \item Validation of solution types
  \item Dashboards with scores
  \item Syntax error highlighting
  \item Suggestions for incorrect predicate names
  \item Warnings about open choice points
\end{itemize}

Students were asked to rate their agreement with the following System Usability Survey Items using a 5-point Likert scale (1 = Strongly Disagree, 5 = Strongly Agree):

\begin{itemize}
  \item I think that I would like to use this system frequently.
  \item I found the system unnecessarily complex.
  \item I thought the system was easy to use.
  \item I think that I would need the support of a technical person to be able to use this system.
  \item I found the various functions in this system were well integrated.
  \item I thought there was too much inconsistency in this system.
  \item I would imagine that most people would learn to use this system very quickly.
  \item I found the system very cumbersome to use.
  \item I felt very confident using the system.
  \item I needed to learn a lot of things before I could get going with this system.
\end{itemize}

Additionally, students rated the usefulness of each feedback component on a scale from 1 (Not useful at all) to 5 (Very useful):

\begin{itemize}
  \item Suggestions for incorrect predicate names
  \item Syntax error highlighting
  \item Dashboards with scores
  \item Select ``Not useful at all'' in this line \textbf{[attention check]}
  \item Automatic test verification
  \item Predicate scoring
  \item Warning for open choice points
  \item Solution type validation (iterative, recursive, functional)
\end{itemize}

\subsection*{Section 4: Role-playing Exercise}

\begin{itemize}
  \item \textbf{Did you solve the role-playing exercise?}
  \begin{itemize}
    \item No
    \item I started but did not finish
    \item Yes
  \end{itemize}

  \item \textbf{Why did you not solve/abandon/finish the exercises?} (Open text response)

  \item \textbf{Did you use LLMs (e.g., ChatGPT, Copilot) to help solve the role-playing exercise?}
  \begin{itemize}
    \item No
    \item Yes
  \end{itemize}

  \item \textbf{Which LLM(s) did you use?} (Open text response)

  \item \textbf{What did you think of the sequence of puzzles from the role-playing exercise on a 5-point scale (1 = Not entertaining at all, 5 = Very entertaining)?}
\end{itemize}

\subsection*{Section 5: Recitation Exercises}

\begin{itemize}
  \item \textbf{Did you solve the recitation exercises using Git/GitLab?}
  \begin{itemize}
    \item No
    \item I started but did not finish
    \item Yes
  \end{itemize}

  \item \textbf{Why did you not solve/abandon/finish the exercises?} (Open text response)

  \item \textbf{Did you use LLMs (e.g., ChatGPT, Copilot) to help with the recitation exercises?}
  \begin{itemize}
    \item No
    \item Yes
  \end{itemize}

  \item \textbf{Which LLM(s) did you use?} (Open text response)
  
\end{itemize}

\subsection*{Section 6: Project}

\begin{itemize}
  \item \textbf{Did you use LLMs (e.g., ChatGPT, Copilot) to help with the project?}
  \begin{itemize}
    \item No
    \item Yes
  \end{itemize}

  \item \textbf{Which LLM(s) did you use?} (Open text response)
\end{itemize}

\subsection*{Section 7: Final Reflections}

\begin{itemize}
    \item \textbf{Please rate the following statements on a 5-point Likert scale (1 = Strongly Disagree, 5 = Strongly Agree)}
    \begin{itemize}
  \item I believe the automated feedback I received on GitLab improved my grade in this course.
  \item I believe the role-playing exercise improved my grade in this course.
  \item Select ``Strongly Agree'' to demonstrate you are paying attention. \textbf{[Attention check]}
  \item I believe the recitation exercises on GitLab improved my grade in this course.
\end{itemize}

    \item \textbf{Please rate the potential utility of proposed features on a 3-point scale (Indifferent, Somewhat Useful, Useful):}
    \begin{itemize}
  \item Suggestions for corrections on incorrect submissions
  \item Showing program output for each test compared to expected output
  \item Automatic identification of bug locations
  \item Detection of infinite loops
\end{itemize}

    \item \textbf{Please suggest any other features you would like to see.} (Open text response)
    \item \textbf{Please leave any other feedback you may have (positive or negative).} (Open text response)
\end{itemize}

\end{document}